\documentclass[numreferences]{crckbked}
\usepackage{epsf}

\newcommand{\eqnref}[1]{Eq.~(\ref{#1})}
\newcommand{\eqnlessref}[1]{(\ref{#1})}

\def\Det{{\rm Det}}

\def\scrD{{\cal D}}
\def\lsim{\mathrel{\lower0.3em\hbox{$\stackrel{\textstyle <}{\sim}$}}}
\def\gsim{\mathrel{\lower0.3em\hbox{$\stackrel{\textstyle >}{\sim}$}}}
\def\scrK{{\cal K}}
\def\negspace{\kern -0.4em}

\def\ehat{{\bf \hat e}}

\def\dvec{\raise 0.3 em \hbox{$^\leftrightarrow$} \kern -0.77 em}

\def\omegahat{\hat%
	{\setbox0=\hbox{$\omega$}%
		\kern-.025em\copy0\kern-\wd0
		\kern.05em\copy0\kern-\wd0
		\kern-.025em\raise.0433em\box0}}

\begin{document}

\begin{opening}
\title{Dual Superconductivity, Effective String Theory, and Regge Trajectories}

\author{M. Baker}\author{R. Steinke}\institute{University of Washington
\\P.O. Box 351560, Seattle, WA 98195, USA}





\begin{abstract}
    We show how an effective field theory of long distance QCD,
    describing a dual superconductor,
    can be expressed as an effective string theory of superconducting
    vortices. We evaluate the semiclassical expansion of this effective
    string theory about a classical rotating string solution
    in any spacetime dimension $D$.
    We show that, after renormalization, the zero point
    energy of the string fluctuations remains finite when the 
    masses of the quarks on the ends of the string approach zero. 
    For $D=26$ the semiclassical energy spectrum of the rotating
    string formally coincides with that of the open string in classical
    Bosonic string theory. However, its physical origin is different.
    It is a semiclassical spectrum of an effective string theory valid
    only for large values of the angular momentum.  For $D=4$ 
    the first semiclassical correction adds the constant $1/12$
    to the classical Regge formula for the angular momentum of
    mesons on the leading Regge trajectory. The excited vibrational 
    modes of the rotating string give rise to daughter Regge 
    trajectories determining the spectrum of hybrid mesons.

\end{abstract}

\end{opening}
\section{The Dual Superconductor Mechanism of Confinement}

In the dual superconductor mechanism of confinement a dual Meissner
effect~\cite{Nambu,Mandelstam,tHooft} confines color electric flux 
to narrow tubes connecting a quark-antiquark pair. The dual Abelian 
Higgs model, which couples dual potentials $C_\mu$ to a monopole Higgs 
field $\phi$, provides an 
example of this mechanism.~\cite{Nielsen+Olesen}
The action $S[C_\mu,\phi]$ of the theory is 
\begin{equation} S[C_\mu,\phi] = \int d^4x \left[ -\frac{\left(\partial_\mu C_\nu - \partial_\nu C_\mu\right)^2}{4} - \frac{\left|\left(\partial_\mu - ig C_\mu\right)\phi \right|^2}{2}
- \frac{\lambda}{4} \left( |\phi|^2 - \phi_0^2 \right)^2 \right] \,.
\label{field action}
\end{equation}
The dual coupling constant is $g=2\pi/e$, where $e$ is
the Yang--Mills coupling constant. In the confined
phase the monopole fields $\phi$ condense to the 
value $\phi_0$. The dual potentials $C_\mu$
acquire a mass $M = g\phi_0$ via a dual Higgs mechanism .
Quarks couple to dual potentials via a Dirac string connecting
the quark-antiquark pair along a line L, the ends of which 
are sources and sinks of color electric flux. The color field of the 
pair destroys the dual Meissner effect near L so that $\phi$
vanishes on L. At distances greater than $1/M$
the monopole field returns to its bulk value $\phi_0$ so that 
 the color field is confined
to a tube of radius $a = 1/M$ surrounding the line L,
which is the location of the
 vortex.~\cite{Baker+Ball+Zachariasen:1991,Nora}  

This model  provides an 
effective theory of long distance QCD,
 i.e. for distances greater than $a$.
Here we want to use this theory to obtain the energy levels
of mesons having large angular momentum.  
Under such circumstances the distance between quarks is much
larger than the flux tube radius and we must take into account
the contribution of flux tube fluctuations to the interaction between
the quark and antiquark,  The fluctuating vortex line L  
sweeps out a world sheet $\tilde x^\mu$  whose  boundary is
the loop $\Gamma$ formed from the world lines  of the moving 
quark-antiquark pair.  Their interaction is determined by 
the Wilson loop $W[\Gamma]$:
\begin{equation}
W[\Gamma] = \int \scrD C_\mu \scrD\phi \scrD\phi^* e^{iS[C_\mu, \phi]} \,.
\label{Wilson loop def}
\end{equation}
The path integral \eqnlessref{Wilson loop def} goes over all field
configurations for which the monopole field  $\phi(x)$
vanishes on some sheet $\tilde x^\mu$ bounded by the loop
$\Gamma$.

\section{The Effective String Theory}
 
The Wilson loop $W[\Gamma]$ describes the quantum fluctuations
of a field theory having classical vortex solutions. We want to
 express the functional integration over 
fields as a path integral over the vortex sheets $\tilde x^\mu$  
to obtain an effective string theory of these vortices. To do this
we carry out the functional integration \eqnlessref{Wilson loop def} 
in two stages:

\begin{enumerate}
\item We integrate over all field configurations
in which the vortex is located on a particular surface
$\tilde x^\mu$, where $\phi(\tilde x^\mu) = 0$. This
integration determines the action $S_{{\rm eff}}[\tilde x^\mu]$
of the effective string theory.
\item We integrate over all
vortex sheets $\tilde x^\mu(\xi)$, $\xi=\xi^a$, $a=1\,,2$. This
integration gives $W[\Gamma]$ the form of an effective 
string theory of vortices.
\end{enumerate}
The action $S_{{\rm eff}}[\tilde x^\mu]$ is invariant under
reparameterizations $\xi \to \xi^\prime(\xi)$ of the world sheet
$\tilde x^\mu(\xi)$ of the vortex. We choose a particular 
parameterization of $\tilde x^\mu$ in terms of the amplitudes
$f^1(\xi)$ and $f^2(\xi)$,
of the two transverse fluctuations of the vortex.
\begin{equation}
\tilde x^\mu(\xi) = x^\mu(f^1(\xi), f^2(\xi), \xi^1, \xi^2) \,.
\label{x param}
\end{equation}
Using the parameterization \eqnlessref{x param}, we can write the
integration  over vortex sheets as a path integral over the 
transverse vortex fluctuations $f^1(\xi)$ and $f^2(\xi)$.
The path integral \eqnlessref{Wilson loop def} over
fields then takes the form:~\cite{Baker+Steinke2}
\begin{equation}
W[\Gamma] = \int \scrD f^1 \scrD f^2 \Delta_{FP} e^{iS_{{\rm eff}}[\tilde x^\mu]} \,,
\label{Wilson loop eff}
\end{equation}
where
\begin{equation}
\Delta_{FP} = \Det\left[ \frac{\epsilon_{\mu\nu\alpha\beta}}{\sqrt{-g}}
\frac{\partial x^\mu}{\partial f^1} \frac{\partial x^\nu}{\partial f^2}
\frac{\partial \tilde x^\alpha}{\partial \xi^1}
\frac{\partial \tilde x^\beta}{\partial \xi^2}
\right]
\end{equation}
is a Faddeev-Popov determinant, and $\sqrt{-g}$ is the square root
of the determinant of the induced metric $g_{ab}$,
\begin{equation}
g_{ab} = \frac{\partial \tilde x^\mu}{\partial \xi^a}
\frac{\partial \tilde x_\mu}{\partial \xi^b} \,.
\end{equation}
The path integral \eqnlessref{Wilson loop eff} goes over 
fluctuations in the shape of the vortex sheet with wave lengths
greater than the radius $1/M$ of the flux tube.
All the long distance fluctuations in $W[\Gamma]$ are contained in
this path integral over string fluctuations. 

The presence of the determinant $\Delta_{FP}$ makes the
path integral \eqnlessref{Wilson loop eff} 
invariant under reparameterizations $\tilde x^\mu(\xi)
\to \tilde x^\mu(\xi(\xi^\prime))$ of the vortex worldsheet. Specifying
the parameterization \eqnlessref{x param} is analogous to fixing
a gauge in a gauge theory, and gives rise to the
 path integral \eqnlessref{Wilson loop eff} over
$f^1(\xi)$ and $f^2(\xi)$ .The factor $\Delta_{FP}$ 
arises from writing the original field theory path 
integral \eqnlessref{Wilson loop def} as a ratio of path integrals 
of two string theories \cite{ACPZ}, and reflects the field theory 
origin of the effective string theory .  This ratio is 
anomaly free.   
Anomalies~\cite{Polyakov:book} present in string theory
are not present in field theory.

\section{The Action of the Effective String Theory}

The parameterization invariant measure in the path 
integral \eqnlessref{Wilson loop eff} is universal and is
independent of the explicit form of the underlying field theory.
On the other hand, the action $S_{{\rm eff}}[\tilde x^\mu]$ 
of the effective string theory is 
not universal. It is determined by the 
integration over all field configurations 
which contain a vortex on the surface $\tilde x^\mu$. However,
for wavelengths $\lambda$ of the string fluctuations greater than 
the flux tube radius $a$, which are those included 
in \eqnlessref{Wilson loop eff}, the action 
$S_{{\rm eff}}[\tilde x^\mu]$ can be expanded in powers of
the extrinsic curvature tensor $\scrK^A_{ab}$ of the sheet
$\tilde x^\mu$,
\begin{equation}
S_{{\rm eff}}[\tilde x^\mu] = - \int d^4x \sqrt{-g} \left[ \sigma
+ \beta \left(\scrK^A_{ab}\right)^2 + ... \right] \,.
\label{curvature expansion}
\end{equation}
The extrinsic curvature tensor is
\begin{equation}
\scrK^A_{ab} = n^A_\mu(\xi) \frac{\partial^2 \tilde x^\mu}
{\partial\xi^a \partial\xi^b} \,,
\end{equation}
where $n^A_\mu(\xi)$, $A = 1,2$ are vectors normal to
the worldsheet at the point $\tilde x^\mu(\xi)$.
The string tension $\sigma$ and the rigidity $\beta$ are
determined by the parameters of the underlying
effective field theory,
whose long distance fluctuations are described by the effective
string theory .

The expansion parameter in \eqnlessref{curvature expansion}
is the ratio $(a/\lambda)^2$ of the square of the 
flux tube radius to the square of 
the wave length of the string fluctuations.
In mesons of angular momentum $J$ this parameter is of order $1/J$.
For mesons of large angular momentum
the effective action \eqnlessref{curvature expansion} can therefore
be approximated by the Nambu--Goto action,
\begin{equation}
S_{{\rm eff}} = S_{{\rm NG}} \equiv -\sigma \int d^4x \sqrt{-g} \,,
\label{Nambu Goto Action}
\end{equation}
and the path integral for $W[\Gamma]$ becomes
\begin{equation}
W[\Gamma] = \int \scrD f^1 \scrD f^2 \Delta_{FP}
e^{i\sigma \int d^4x \sqrt{-g}} \,.
\label{Wilson Nambu Goto}
\end{equation}
In the next section we describe the 
results~\cite{Baker+Steinke2,ron}
of a semiclassical expansion 
of this effective string theory .

\section{The Semiclassical Calculation in the Background of a Rotating String}

\begin {figure}[H]
    \begin{center}
	\null \hfill \epsfbox{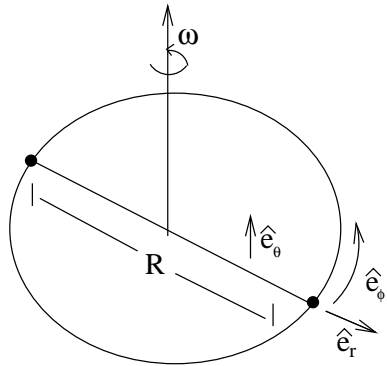} \hfill \null
    \end{center}
    \caption{The string coordinate system}
    \label{rot string fig}
\end {figure}

We calculate $W[\Gamma]$ in the leading semiclassical approximation
in the background of a worldsheet generated by a straight string
attached to quarks rotating with uniform angular velocity $\omega$
(See Fig.~\ref{rot string fig}).
The quarks have masses $m_1$ and $m_2$,
move with velocities $v_1 = \omega R_1$ and $v_2 = \omega R_2$,
and are separated by a fixed distance $R = R_1 + R_2$. The
parameters $\xi = (t, r)$ are the time $t$ and the coordinate $r$,
which runs along the string from $-R_1$ to $R_2$, so that the
transverse velocity of the straight string is zero when $r=0$.

The amplitudes $f^1(\xi)$ and $f^2(\xi)$ of the transverse fluctuations are
the spherical coordinates $\theta(r,t)$ and $\phi(r,t)$
of a point on the string. These angles are defined in
an unconventional manner so that $\theta(r,t) = \phi(r,t) = 0$
is a straight string rotating in the $xy$ plane. The angle $\theta(r,t)$
is the fluctuation perpendicular to the plane of rotation, and
the angle $\phi(r,t)$ is the fluctuation lying in the plane
of the rotating string.  

The ends of the string are fixed to their classical trajectories,
\begin{equation}
\theta(-R_1, t) = \theta(R_2, t) = \phi(-R_1, t) = \phi(R_2, t) = 0 \,.
\label{classical boundary angles}
\end{equation}
The fluctuating worldsheet $\tilde x^\mu$ then has the parameterization
$\tilde x^\mu(r,t)$ given by
\begin{eqnarray}
\tilde x^\mu(r,t) &=&  x^\mu(\theta(r, t), \phi(r, t), r, t)
\nonumber \\
&=& t \ehat_0^\mu + r \Big[ \cos\theta(r,t) 
\cos\left(\phi(r,t) + \omega t\right)
\ehat_1^\mu \nonumber\\& & + \cos\theta(r,t) \sin\left(\phi(r,t) + \omega t\right) \ehat_2^\mu
 - \sin\theta(r,t) \ehat_3^\mu \Big] \,,
\label{x mu param}
\end{eqnarray}
where $\ehat_\alpha^\mu$, $\alpha = 0\,,...\,3$ are unit vectors along the
four fixed spacetime axes, $\ehat_\alpha^\mu = \delta_\alpha^\mu$.
The parameterization \eqnlessref{x mu param} is a concrete example
of \eqnlessref{x param}.

The classical rotating straight string $\bar x^\mu(r,t)$ has the
parameterization, 
\begin{eqnarray}
\bar x^\mu(r,t) &=& x^\mu\left( \theta(r,t) = 0, \phi(r,t) = 0, r, t \right)
\nonumber \\
&=& t \ehat_0^\mu + r \left[ \cos\omega t \ehat_1^\mu + \sin\omega t \ehat_2^\mu
\right] \,.
\end{eqnarray}
The corresponding metric $\bar g_{ab} = g_{ab}[\bar x^\mu]$
and classical action $S_{{\rm NG}}[\bar x^\mu]$ are independent of
the time $t$, so that $W[\Gamma]$ has the form
\begin{equation}
W[\Gamma] = e^{iTL^{{\rm string}}(R_1, R_2, \omega)} \,,
\label{Wilson loop integrated}
\end{equation}
where $T$ is the elapsed time. For massless quarks, the ends of the
string move with the velocity of light, and singularities
appear in $L^{{\rm string}}$. To regulate these singularities,
we retain the quark mass as a cutoff and take the massless limit
at the end when evaluating physical quantities.

The Lagrangian $L^{{\rm string}}$ is the sum of a classical part
$L^{{\rm string}}_{{\rm cl}}$ and a fluctuating part 
$L^{{\rm string}}_{{\rm fluc}}$,
\begin{equation}
L^{{\rm string}} = L^{{\rm string}}_{{\rm cl}} + L^{{\rm string}}_{{\rm fluc}} \,,
\label{L string separated}
\end{equation}
where
\begin{equation}
L^{{\rm string}}_{{\rm cl}} = -\frac{\sigma}{T} \int d^2\xi \sqrt{-\bar g}
= -\sigma \int_{-R_1}^{R_2} dr \sqrt{1 - r^2 \omega^2} \,.
\end{equation}
We will calculate $L^{{\rm string}}_{{\rm fluc}}$ in $D$
spacetime
dimensions.The fluctuation  $\theta(r,t)$ is replaced by 
$D-3$ fluctuations
perpendicular to the plane of rotation and there is still just 
1 fluctuation 
in the plane of rotation.

The effective Lagrangian for the quark--antiquark pair is obtained 
by adding
quark mass terms to $L^{{\rm string}}$:
\begin{equation}
L_{{\rm eff}}(R_1, R_2, \omega) = - \sum_{i=1}^2 m_i \sqrt{1-v_i^2}
+ L^{{\rm string}}(R_1, R_2, \omega) \,.
\end{equation}
The effective Lagrangian is the sum of a classical part 
and a fluctuating part,
\begin{equation}
L_{{\rm eff}}(R_1, R_2, \omega) = L_{{\rm cl}}(R_1, R_2, \omega)
+ L^{{\rm string}}_{{\rm fluc}}(R_1, R_2, \omega) \,,
\end{equation}
where
\begin{eqnarray}
L_{{\rm cl}} &=& - \sum_{i=1}^2 m_i \sqrt{1-v_i^2}
 - \sigma \int_{-R_1}^{R_2} dr \sqrt{1 - r^2 \omega^2}
\nonumber \\
&=& -\sum_{i=1}^2 \left[m_i \gamma_i^{-1} + \sigma \frac{R_i}{2} \left( \frac{\arcsin(v_i)}{v_i}
+ \gamma_i^{-1} \right) \right] \,,
\label{L_cl def formal}
\end{eqnarray}
with
\begin{equation}
\gamma_i^{-1} = \sqrt{1-v_i^2} \,, \kern 1 in v_i = \omega R_i \,.
\end{equation}

The expression for $L^{{\rm string}}_{{\rm fluc}}$ is obtained
from \eqnlessref{Wilson loop integrated} and the semiclassical
calculation of $W[\Gamma]$. It contains terms which are quadratically,
linearly, and logarithmically divergent in the cutoff $M$.
The quadratically divergent term is a renormalization of the
string tension, the linearly divergent term is a renormalization
of the quark mass, and the logarithmically divergent term is
proportional to the integral of the scalar curvature
over the whole worldsheet~\cite{Luscher1}. After absorbing the
quadratically and linearly divergent terms into renormalizations,
we obtain \cite{Baker+Steinke2, ron}   
\begin{equation}
L^{{\rm string}}_{{\rm fluc}}(R_1, R_2, \omega)
= \frac{\pi (D-2)}{24R_p} - \sum_{i=1}^2 \frac{\omega v_i \gamma_i}{\pi}
\left[ \ln\left(\frac{M R_i}{\gamma_i^2 - 1}\right) + 1 \right]
+ \frac{\omega}{2} + \omega f(v_1, v_2) \,,
\label{L fluc eval}
\end{equation}
where $R_p$ is the proper length of the string,
\begin{equation}
R_p = \frac{1}{\omega} \left( \arcsin v_1 + \arcsin v_2 \right) \,,
\label{proper length}
\end{equation}
and
\begin{equation}
f(v_1, v_2) = -\frac{1}{\pi} \int_0^\infty ds \ln\left[
\frac{s^2 + (v_1\gamma_1 + v_2\gamma_2) s \coth(s R_p \omega)
+ v_1 \gamma_1 v_2 \gamma_2}{(s + v_1\gamma_1)(s + v_2\gamma_2)} \right] \,.
\label{f v_1 v_2 def}
\end{equation}
The function $f(v_1,v_2)$ vanishes when $v_1$ and $v_2$ approach
unity, so that the last term in \eqnlessref{L fluc eval} is
small for relativistic quarks. The contribution to
$L^{{\rm string}}_{{\rm fluc}}$ of the additional
$D-4$ fluctuations perpendicular to the plane of rotation, coming from
the extension to D dimensional spacetime, is contained in the first term
in \eqnlessref{L fluc eval}.

In the limit $\omega \to 0$, $R_p \to R_1 + R_2 = R$, and
$L^{{\rm string}}_{{\rm fluc}}$ reduces to the result of L\"uscher
\cite{Luscher2}
for the correction to the static quark--antiquark potential
due to string fluctuations,
\begin{equation}
V_{\hbox{\scriptsize L\"uscher}} = - L_{{\rm fluc}}^{{\rm string}}
(R_1, R_2, \omega = 0) = - \frac{\pi (D-2)}{24R} \,.
\end{equation}
For $\omega \ne 0$,
$L^{{\rm string}}_{{\rm fluc}}$ contains a logarithmically divergent
part. We simplify this term using the classical equation of
motion,
\begin{equation}
\frac{\partial L_{{\rm cl}}}{\partial R_i} \Big|_{R_i = \bar R_i} = 0 \,,
\label{canonical classical eqn}
\end{equation}
to express $\bar R_i$ in terms of $\omega$. \eqnref{canonical classical eqn}
gives the relation
\begin{equation}
\sigma \bar R_i = m_i \left(\bar \gamma_i^2 - 1\right) \,,
\label{classical eqn}
\end{equation}
where $\bar\gamma_i$ is equal to $\gamma_i$ evaluated at $R_i = \bar R_i$.
The solution of \eqnlessref{classical eqn} for $\bar R_i$ as a function of $\omega$ is
\begin{equation}
\bar R_i = \frac{1}{\omega} \left(\sqrt{\left(\frac{m_i\omega}{2\sigma}\right)^2
+ 1} - \frac{m_i\omega}{2\sigma}\right) \,.
\label{solve for R_i}
\end{equation}
Using the relation \eqnlessref{classical eqn} in \eqnlessref{L fluc eval}
gives
\begin{equation}
L^{{\rm string}}_{{\rm fluc}}(\omega) = \frac{\pi(D-2)}{24 R_p} - \sum_{i=1}^2
\frac{\omega v_i \bar\gamma_i}{\pi} \left[ \ln\left(\frac{Mm_i}{\sigma}\right) + 1 \right]
+ \frac{\omega}{2} + \omega f(v_1, v_2) \,.
\label{L fluc pre renorm}
\end{equation}
Use of the classical equations of motion has eliminated the logarithmic
dependence of $L^{{\rm string}}_{{\rm fluc}}$ on the dynamical 
parameter $\omega$ .

The logarithmically divergent term in \eqnlessref{L fluc pre renorm} 
is proportional to the quantity $\omega v_i \bar \gamma_i$ which
diverges when $m_i \to 0$. This term can be absorbed into a 
renormalization of a term in the boundary action called the
geodesic curvature \cite{Alvarez:1983} so that the theory will be finite in the $m_i \to 0$
limit.

\section{Renormalization of the Geodesic Curvature}

\label{geodesic curvature section}

For a straight string rotating
with angular velocity $\omega$, the geodesic curvature is equal
to $\omega v_i\gamma_i^2$.  The logarithmic divergence
in \eqnlessref{L fluc pre renorm}
can then be removed by adding to the quark mass term in the boundary
Lagrangian a counterterm  containing the geodesic
curvature. This gives
\begin{eqnarray}
L^{{\rm boundary}} &=& -\sum_{i=1}^2 \gamma_i^{-1} \left[ m_i
+ \kappa_i \omega v_i \gamma_i^2  \right] \,,
\label{L boundary expand}
\end{eqnarray}
where the second term in \eqnlessref{L boundary expand} is  
the geodesic curvature term.
The logarithmic divergence
in \eqnlessref{L fluc pre renorm} can then be regarded 
as a renormalization of the coefficient $\kappa_i$  
of the geodesic curvature
in \eqnlessref{L boundary expand}. 

Since the quantity $\omega v_i \bar \gamma_i$
diverges when $m_i \to 0$, 
the requirement that the theory is finite
in the $m_i \to 0$ limit forces the 
renormalized value of $\kappa_i$ to be zero.
(We take the $m_i \to 0$ limit with
the cutoff $M$ fixed, because we have an effective theory).
Removing the terms in \eqnlessref{L fluc pre renorm} proportional
to $\omega v_i \gamma_i$ gives an expression 
for $L^{{\rm string}}_{{\rm fluc}}$
which is applicable in the massless quark limit,
\begin{equation}
L_{{\rm fluc}}^{{\rm string}} = \frac{\pi(D-2)}{24R_p} + \frac{\omega}{2} 
+ \omega f(v_1, v_2) \,.
\label{L fluc renorm}
\end{equation}
In the case of two light quarks,  $m_1 = m_2 = 0$
($\bar\gamma_1, \bar\gamma_2 \to \infty$), Eqs.~\eqnlessref{proper length}
and \eqnlessref{f v_1 v_2 def} give $R_p = \pi/\omega$ and
$f(v_1, v_2) = 0$, so that \eqnlessref{L fluc renorm} becomes
\begin{equation}
L^{{\rm string}}_{{\rm fluc}}(\omega) \Big|_{m_1 = m_2 = 0}
= \frac{\omega (D-2)}{24} + \frac{\omega }{2} \,.
\label{L fluc ll}
\end{equation}
The first term in \eqnlessref{L fluc ll} is the negative of L\"uscher
potential with the length $R$ of the string replaced by its proper length 
$R_p = \pi/\omega$. It is the contribution of $D-2$ tranverse fluctuations
in the background of a flat metric. The $\frac{\omega}{2}$ term accounts
for the curvature of the classical background metric 
$\bar g_{ab} = g_{ab}[\bar x^\mu]$ generated by the rotating string.

In the case of one heavy and one light quark, $m_1 \to \infty$ ($v_1 \to 0$)
and $m_2 = 0$ ($\bar\gamma_2 \to \infty$), $R_p = \pi/2\omega$ and
\begin{equation}
f(v_1 = 0, v_2 \to 1) = -\frac{1}{\pi} \int_0^\infty \ln\coth\left(\frac{\pi}{2} s\right)
= -\frac{\omega}{4} \,,
\end{equation}
so
\begin{equation}
L^{{\rm string}}_{{\rm fluc}}(\omega) \Big|_{{m_1 \to \infty} \atop {m_2 = 0}}
= \frac{\omega (D-2)}{12} + \frac{\omega}{4} \,.
\label{L fluc hl}
\end{equation}
In this talk we use \eqnlessref{L fluc ll}
to determine Regge trajectories of mesons
composed of two light quarks.

\section{Fluctuations in the Motion of the Quarks at the Ends of the String}

\label{end motion section}

In the previous discussion, the quark--antiquark pair moved in
a fixed classical trajectory in the $xy$ plane 
(See Fig.~\ref{rot string fig}
and Eq's \eqnlessref{classical boundary angles} 
and \eqnlessref{solve for R_i}).
To take into account the fluctuations of the 
positions ${\bf \vec x}_1(t)$
and ${\bf \vec x}_2(t)$ of the quarks at the ends of the rotating 
string, we extend the functional
integral \eqnlessref{Wilson Nambu Goto} to
include a path integral over ${\bf \vec x}_1(t)$ and
${\bf\vec x}_2(t)$, and add the action of the quarks to the
string action \eqnlessref{Nambu Goto Action}. This
extension replaces $W[\Gamma]$ by the partition function $Z$,
\begin{equation}
Z = \frac{1}{Z_b} \int \scrD f^1(\xi) \scrD f^2(\xi) \scrD {\bf\vec x}_1(t) 
\scrD {\bf\vec x}_2(t) \Delta_{FP} e^{-i\sigma \int d^2\xi \sqrt{-g}
+iS_{{\rm quark}}} \,,
\label{string partition}
\end{equation}
where $S_{{\rm quark}}$ is the action of two free (scalar) quarks,
\begin{equation}
S_{{\rm quark}} = -  \sum_{i=1}^2 m_i \int_{-T/2}^{T/2} dt 
\sqrt{1 - \dot{\vec x}_i^2(t)} \,,
\label{quark action}
\end{equation}
and where $Z_b$ is the corresponding quark partition function ,
\begin{equation}
Z_b = \int \scrD {\bf\vec x}_1(t)\scrD {\bf\vec x}_2(t)
e^{iS_{{\rm quark}}} \,.
\label{quark partition function}
\end{equation}
Dividing by $Z_b$ removes the vacuum energy of the quarks.

We use the methods of  Dashen, Hasslacher, Neveu ~\cite {DHN} to carry
out a semiclassical calculation of the partition function $Z$ 
around periodic classical solutions and determine the energies
of the physical meson states. We find :
\begin{enumerate}
\item For massless quarks, the contribution to $L_{{\rm eff}}(\omega)$
arising from the fluctuations in the
motion of the ends of the string 
along the vector
$\hat e_r$ shown in Fig.~\ref{rot string fig} 
vanishes. For massless quarks the longitudinal mode is unphysical
and the meson energy comes from the fluctuations of the interior of the 
string, i.e. $L_{{\rm fluc}}(\omega) = L^{{\rm string}}_{{\rm fluc}}(\omega)$.

(For massive quarks the fluctuations of the boundary also contribute
to the meson energy.)
\item The usual WKB quantization condition for angular momentum.
\end{enumerate}

The angular momentum $J$,
\begin{equation}
J = \frac{d L_{{\rm eff}}(\omega)}{d\omega} \,,
\end{equation}
takes on the values,
\begin{equation}
J = l + \frac{1}{2} \kern 1 in l = 0, 1, 2,... \,.
\label{J quant}
\end{equation}
The energy $E(\omega)$ is given by the corresponding Hamiltonian,
\begin{equation}
E(\omega) = \omega \frac{d L_{{\rm eff}}(\omega)}{d\omega}
- L_{{\rm eff}}(\omega) \,.
\end{equation}

Eq's \eqnlessref{L_cl def formal} and \eqnlessref{L fluc ll} give
 $L_{{\rm fluc}}(\omega) / L_{{\rm cl}}(\omega) 
\sim \omega^2/\sigma \sim 1/J$
so that for large $J$, $L_{{\rm
fluc}}(\omega)$ can be treated as a perturbation. The meson energy
is then
\begin{equation}
E(\omega) = E_{{\rm cl}}(\omega) - L_{{\rm fluc}}(\omega) \,,
\label{E of J}
\end{equation}
where
\begin{equation}
E_{{\rm cl}}(\omega) = \omega \frac{d L_{{\rm cl}}(\omega)}{d\omega}
- L_{{\rm cl}}(\omega) \,,
\label{E class}
\end{equation}
and where $\omega$ is given
as a function of $J$ by the classical relation
\begin{equation}
J = \frac{d L_{{\rm cl}}(\omega)}{d\bar\omega} \,.
\label{J class}
\end{equation}
Eqs. \eqnlessref{J class} and \eqnlessref{E of J} determine 
meson Regge trajectories in the 
leading semiclassical  approximation.
\section{Meson Regge Trajectories}

We first evaluate 
$L_{{\rm cl}}(\omega)$
to obtain classical Regge trajectories. For massless quarks Eqs.
\eqnlessref{L_cl def formal}, \eqnlessref{E class}, and 
\eqnlessref{J class} give 
\begin{equation}
L_{{\rm cl}}(\omega) = - \frac{\pi\sigma}{2\omega} \,, \kern 0.5in
J = \frac{\pi\sigma}{2\omega^2} \,, \kern 0.5in
E_{{\rm cl}} = \frac{\pi\sigma}{\omega} \,, \kern 0.5in
J = \frac{E_{{\rm cl}}^2}{2\pi\sigma} \,.
\label{classical Regge eqn ll}
\end{equation}

Using Eqs. \eqnlessref{E of J},
\eqnlessref{L fluc ll}, and \eqnlessref{classical Regge eqn ll} 
to include the correction to the energy 
due to the zero point energy of the string fluctuations gives
\begin{equation}
E(\omega) = E_{cl}(\omega) - L_{{\rm fluc}}(\omega)
= \frac{\pi\sigma}{\omega} - \frac{D-2}{24} \omega -\frac{\omega}{2} \,.
\label{E of omega ll}
\end{equation}

The energies $E_n(\omega)$ of the excited states of the rotating string
(light hybrid mesons) are
obtained by adding  to
\eqnlessref{E of omega ll} the energies $k\omega$ of the excited
vibrational modes.
\begin{equation}
E_n(\omega) = \frac{\pi\sigma}{\omega} - \frac{D-2}{24} \omega
-\frac{\omega}{2} + n \omega \,.
\label{hybrid energy ll}
\end{equation}
Since there are many combinations of string normal modes
which give the same $n$ (e.g., a doubly excited $k=1$ mode
and a singly excited $k=2$ mode each give $n=2$), the spectrum
is highly degenerate. There are $D-2$  trajectories having $n=1$,
each corresponding to a single excitation of one of
the $k=1$ normal modes. Higher values of $n$ have higher
degeneracies.

 The value of $\omega$ is given as a function of $J$ through the
classical relation $\omega = \sqrt{\pi\sigma/2J}$. Squaring
both sides of \eqnlessref{hybrid energy ll} yields
\begin{eqnarray}
E_n^2 & = & 2\pi\sigma\left(J - \frac{D-2}{24} - \frac{1}{2} +
 n + O\left( \frac{n^2}{J}\right)\right) 
\nonumber \\
&=& 2\pi\sigma\left(l + n - \frac{D-2}{24} + O\left(
\frac{n^2}{l}\right)\right) \, ,
\label{E^2 l}
\end{eqnarray}
where we have used the WKB quantization condition \eqnlessref{J quant} .
Setting $D=4$ in \eqnlessref{E^2 l}  and  solving for $l$ we obtain the
the  Regge trajectories,
\begin{equation}
l = \frac{E^2}{2\pi\sigma} + \frac{1}{12} - n
 \kern 1 in n =0, 1, 2, ... \,.
\label{hybrid l of E ll}
\end{equation}
The excited states, $n>0$,  of the rotating
string give rise to daughter Regge trajetories determining
the spectrum of hybrid mesons composed of zero mass quarks.
The first semiclassical correction to the leading
Regge trajectory, $n=0$, adds the constant $1/12$ to the
classical Regge formula. The small size of this correction
could explain why Regge trajectories are linear at values of $l$
of order 1.

\section{Comparison with Bosonic String Theory}

For $D=26$, Eq. \eqnlessref{E^2 l}
yields the spectrum
\begin{equation}
E^2 = 2\pi\sigma\left(l+n - 1 
+O\left({n^2}/l \right)\right) \, .
\label{26 dim spectrum}
\end{equation}
The spectrum of energies \eqnlessref{26 dim spectrum}
coincides with
the spectrum of open strings  in classical bosonic string theory. 
However, \eqnlessref{26 dim spectrum}
is valid only in the leading semiclassical approximation, 
so that it cannot be used for $l=0$, where it would yield the scalar tachyon
of the open bosonic string.


\section{Summary and Conclusions}

\begin{enumerate}
\item We have seen how the dual superconducting 
model of confinement leads to an effective string theory of 
long distance QCD. We have calculated, in the semiclassical
large angular momentum 
 domain , where 
this theory is applicable, 
the effect of string fluctuations on Regge trajectories
for mesons containing light quarks. Similar calculations
~\cite{ron} give Regge 
\nopagebreak 
trajectories for mesons
containing one heavy and one light quark.
\item The spectrum of the energies of the excited states
formally coincides with the spectrum of the open
string of Bosonic string theory in its critical dimension
$D = 26$. Here, we obtained the  spectrum 
for any $D$, but only for large $l$, from the semiclassical 
expansion of an effective string theory.  The presence of
functional determinant $\Delta_{FP}$ 
in the path integral \eqnlessref{Wilson Nambu Goto}
was essential in obtaining this result.
\item We treated the light quarks as massless scalar
particles to determinine the
energies of  the high angular momentum
excited states of mesons. However,
the effect of chiral symmetry breaking, generating a 
constituent quark mass,
must play a dominant role in determining the masses of mesons
which are ground states of quark--antiquark systems.
\item The derivation of the effective string theory
made no use of the details of the effective  field theory
from which it was obtained. Furthermore, 
arguments based on the work of 't Hooft~\cite{tHooft2} indicate that  the confined phase
of a non Abelian gauge
theory is characterized by a dual order parameter, which
vanishes in regions  of space where "dual supercondutivity" 
is destroyed. This generic description of the confined phase of QCD
leads to the effective string theory of long distance
QCD described here, which provides a concrete
picture of the QCD string.
\end{enumerate}

\section*{Acknowledgements}

We would like to thank S.Olejnik and the other organizers
for providing the opportunity to participate in this stimulating
workshop.



\end{document}